\documentclass[submission]{eptcs}
\usepackage[cp1250]{inputenc}
\usepackage[T1]{fontenc}
\usepackage[english]{babel}
\usepackage{amssymb,amsmath,amsfonts}
\usepackage{amsthm}

\newtheorem{lem}{Lemma}
\newtheorem{thm}[lem]{Theorem}
\newtheorem{prop}[lem]{Proposition}

\newtheorem{de}[lem]{Definition}
\newtheorem{ex}[lem]{Example}
\newtheorem{pozn}[lem]{Remark}

\providecommand{\event}{WORDS 2011} 
\usepackage{breakurl}             

\title{Substitutions over infinite alphabet generating $(-\beta)$-integers}
\author{Daniel Dombek
\institute{Department of Mathematics FNSPE}
\institute{Czech Technical University in Prague\\ Czech Republic}
\email{dombedan@fjfi.cvut.cz}}

\def\titlerunning{Substitutions over infinite alphabet generating $(-\beta)$-integers}
\def\authorrunning{Daniel Dombek}

\begin{document}
\maketitle

\section{Introduction}

This contribution is devoted to the study of positional numeration systems with negative base introduced by Ito and Sadahiro in 2009, called $(-\beta)$-expansions. We give an admissibility criterion for more general case of $(-\beta)$-expansions and discuss the properties of the set of $(-\beta)$-integers, denoted by $\mathbb{Z}_{-\beta}$. We give a description of distances within $\mathbb{Z}_{-\beta}$ and show that this set can be coded by an infinite word over an infinite alphabet, which is a fixed point of a non-erasing non-trivial morphism.

\section{Numeration with negative base}

In 1957, R\'enyi introduced positional numeration system with positive real base $\beta>1$ (see \cite{Renyi}).The $\beta$-expansion of $x\in[0,1)$ is defined as the digit string $d_\beta(x)=0\bullet x_1x_2x_3\cdots$, where \[x_i=\lfloor\beta T_\beta^{i-1}(x)\rfloor\quad\text{and}\quad T_\beta(x)=\beta x-\lfloor\beta x\rfloor\,.\] It holds that \[x=\frac{x_1}{\beta}+\frac{x_2}{\beta^2}+\frac{x_3}{\beta^3}+\cdots\,.\] Note that this definition can be naturally extended so that any real number has a unique $\beta$-expansion, which is usually denoted $d_\beta(x)=x_k x_{k-1}\cdots x_1x_0\bullet x_{-1}x_{-2}\cdots$, where $\bullet$, the fractional point, separates negative and non-negative powers of $\beta$. In analogy with standard integer base, the set $\mathbb{Z}_\beta$ of $\beta$-integers is defined as the set of real numbers having the $\beta$-expansion of the form $d_\beta(x)=x_k x_{k-1}\cdots x_1x_0\bullet 0^\omega$.

$(-\beta)$-expansions, a numeration system built in analogy with R\'enyi $\beta$-expansions, was introduced in 2009 by Ito and Sadahiro (see \cite{ItoSadahiro}). They gave a lexicographic criterion for deciding whether some digit string is the $(-\beta)$-expansion of some $x$ and also described several properties of $(-\beta)$-expansions concerning symbolic dynamics and ergodic theory. Note that dynamical properties of $(-\beta)$-expansions were also studied by Frougny and Lai (see \cite{ChiaraFrougny}). We take the liberty of defining $(-\beta)$-expansions in a more general way, while an analogy with positive base numeration can still be easily seen.

\begin{de}\label{de:minusbeta}
Let $-\beta<-1$ be a base and consider $x\in[l,l+1)$, where $l\in\mathbb{R}$ is arbitrary fixed. We define the $(-\beta)$-expansion of $x$ as the digit string $d(x)=x_1x_2x_3\cdots$, with digits $x_i$ given by 
\begin{equation}\label{predpis_cifry}
x_i=\lfloor-\beta T^{i-1}(x)-l\rfloor\,,
\end{equation} where $T(x)$ stands for the generalised $(-\beta)$-transformation \begin{equation}\label{transformace}
T:[l,l+1)\rightarrow[l,l+1)\,,\quad T(x)=-\beta x-\lfloor-\beta x-l\rfloor\,.
\end{equation}
\end{de}

\noindent It holds that \[x=\frac{x_1}{-\beta}+\frac{x_2}{(-\beta)^2}+\frac{x_3}{(-\beta)^3}+\cdots\] and the fractional point is again used in the notation, $d(x)=0\bullet x_1x_2x_3\cdots$.

The set of digits used in $(-\beta)$-expansions of numbers (in the latter referred to as the alphabet of $(-\beta)$-expansions) depends on the choice of $l$ and can be calculated directly from (\ref{predpis_cifry}) as \begin{equation}\label{abeceda}
\mathcal{A}_{-\beta,l}=\big\{\lfloor-l(\beta+1)-\beta\rfloor,\ldots,\lfloor-l(\beta+1)\rfloor\big\}\,.
\end{equation}

We may demand that the numeration system possesses various properties. Let us summarise the most natural ones:

\begin{itemize}
\item The most common requirement is that zero is an allowed digit. We see that $0\in\mathcal{A}_{-\beta,l}$ is equivalent to $0\in[l,l+1)$ and consequently $l\in(-1,0]$. Note that this implies $d(0)=0\bullet 0^\omega$.
\item We may require that $\mathcal{A}_{-\beta,l}=\{0,1,\ldots,\lfloor\beta\rfloor\}$. This is equivalent to the choice $l\in\big(-\frac{\lfloor\beta\rfloor+1}{\beta+1},-\frac{\beta}{\beta+1}\big]$.
\item So far, $(-\beta)$-expansions were defined only for numbers from $[l,l+1)$. In R\'enyi numeration, the $\beta$-expansion of arbitrary $x\in\mathbb{R}^+$ (expansions of negative numbers differ only by ``$-$'' sign) is defined as $d_\beta(x)=x_kx_{k-1}\cdots x_1x_0\bullet x_{-1}x_{-2}\cdots$, where $k\in\mathbb{N}$ satisfies $\frac{x}{\beta^k}\in[l,l+1)$ and $d_\beta\big(\frac{x}{\beta^k}\big)=0\bullet x_kx_{k-1}x_{k-2}\cdots$. The same procedure does not work for $(-\beta)$-expansions in general. A necessary and sufficient condition for the existence of unique $d(x)$ for all $x\in\mathbb{R}$ is that $-\frac{1}{\beta}[l,l+1)\subset[l,l+1)$. This is equivalent to the choice $l\in\big(-\frac{\beta}{\beta+1},-\frac{1}{\beta+1}\big]$. Note that this choice is disjoint with the previous one, so one cannot have uniqueness of $(-\beta)$-expansions and non-negative digits bounded by $\beta$ at the same time.
\end{itemize}

Let us stress that in the following we will need $0$ to be a valid digit. Therefore, we shall always assume $l\in(-1,0]$. Note that we may easily derive that the digits in the alphabet $\mathcal{A}_{-\beta,l}$ are then bounded by $\lceil\beta\rceil$ in modulus.

\section{Admissibility}

In R\'enyi numeration there is a natural correspondence between ordering on real numbers and lexicographic ordering on their $\beta$-expansions. In $(-\beta)$-expansions, standard lexicographic ordering is not suitable anymore, hence a different ordering on digit strings is needed.

The so-called alternate order was used in the admissibility condition by Ito and Sadahiro and it will work also in the general case. Let us recall the definition. For the strings \[u,v\in(\mathcal{A}_{-\beta,l})^\mathbb{N}\,,\quad u=u_1u_2u_3\cdots\quad\hbox{and}\quad v=v_1v_2v_3\cdots\] we say that $u\prec_{alt} v$ ($u$ is less than $v$ in the alternate order) if $u_m(-1)^m<v_m(-1)^m$, where $m=\min\{k\in\mathbb{N}\ \mid\ u_k\neq v_k\}$. Note that standard ordering between reals in $[l,l+1)$ corresponds to the alternate order on their respective $(-\beta)$-expansions.

\begin{de}
An infinite string $x_1x_2x_3\cdots$ of integers is called $(-\beta)$-admissible (or just admissible), if there exists an $x\in[l,l+1)$ such that
$x_1x_2x_3\cdots$ is its $(-\beta)$-expansion, i.e. $x_1x_2x_3\cdots=d(x)$.
\end{de}

We give the criterion for $(-\beta)$-admissibility (proven in $\cite{DMP}$) in a form similar to both Parry lexicographic condition (see \cite{Parry}) and Ito-Sadahiro admissibility criterion (see \cite{ItoSadahiro}).

\begin{thm}\emph{(\cite{DMP})}\label{thm:hlavni}
An infinite string $x_1x_2x_3\cdots$ of integers is $(-\beta)$-admissible, if and only if
\begin{equation}\label{eq:admis}
l_1l_2l_3\cdots\preceq_{alt} x_ix_{i+1}x_{i+2}\cdots\prec_{alt} r_1r_2r_3\cdots\,,
\qquad\hbox{for all $\ i\geq 1$},
\end{equation} where $l_1l_2l_3\cdots=d(l)$ and $r_1r_2r_3\cdots=d^*(l+1)=\lim_{\epsilon\to 0+}d(l+1-\epsilon)$. 
\end{thm}

\begin{pozn}\label{pozn:IS}
Ito and Sadahiro have described the admissibility condition for their numeration system considered with $l=-\frac{\beta}{\beta+1}$. This choice imply for any $\beta$ the alphabet of the form $\mathcal{A}_{-\beta,l}=\{0,1,\ldots,\lfloor\beta\rfloor\}$. They have shown that in this case the reference strings used in the condition in Theorem~\ref{thm:hlavni} (i.e. $d(l)=l_1l_2l_3\cdots$ and $d^*(l+1)=r_1r_2r_3\cdots$) are related in the following way: \[r_1r_2r_3\cdots = 0l_1l_2l_3\cdots\]
if $d(l)$ is not purely periodic with odd period length, and, \[r_1r_2r_3\cdots = \big(0l_1l_2\cdots l_{q-1}(l_{q}-1)\big)^\omega\,,\] if $d(l) = \big(l_1l_2\cdots l_{q}\big)^\omega$, where $q$ is odd.
\end{pozn}

\begin{pozn}\label{pozn:bal}
Besides Ito-Sadahiro case and the general one, we may consider another interesting example, the choice $l=-\frac{1}{2}$, $\beta\notin 2\mathbb{Z}+1$. This leads to a numeration defined on ``almost symmetric'' interval $[-\frac{1}{2},\frac{1}{2})$ with symmetric alphabet \[\mathcal{A}_{-\beta,-\frac{1}{2}}= \Big\{\overline{\Big\lfloor\frac{\beta+1}{2}\Big\rfloor}, \ldots, \overline{1}, 0, 1, \ldots \Big\lfloor\frac{\beta+1}{2}\Big\rfloor\Big\}\,.\] 

Note that we use the notation $(-a)=\overline{a}$ for shorter writing of negative digits. If we denote the reference strings as usual, i.e. $d\big(-\frac{1}{2}\big)=l_1l_2l_3\cdots$ and $d^*\big(\frac{1}{2}\big)=r_1r_2r_3\cdots$, the following relation can be shown: \[r_1r_2r_3\cdots = \overline{l_1l_2l_3\cdots}\]
if $d(l)$ is not purely periodic with odd period length, and, \[r_1r_2r_3\cdots = \big(\overline{l_1l_2\cdots l_{q-1}(l_{q}-1)}l_1l_2\cdots l_{q-1}(l_{q}-1)\big)^\omega\,,\] if $d(l) = \big(l_1l_2\cdots l_{q}\big)^\omega$, where $q$ is odd. 
\end{pozn}

\section{$(-\beta)$-integers}

We have already discussed basic properties of $(-\beta)$-expansions and the question of admissibility of digit strings. In the following, $(-\beta)$-admissibility will be used to define the set of $(-\beta)$-integers.

Let us define a ``value function'' $\gamma$. Consider a finite digit string $x_{k-1}\cdots x_1 x_0$, then $\gamma(x_{k-1},\cdots x_1 x_0)=\sum_{i=0}^{k-1}x_i(-\beta)^i$.

\begin{de}\label{de:integers}
We call $x\in\mathbb{R}$ a $(-\beta)$-integer, if there exists a $(-\beta)$-admissible digit string $x_kx_{k-1}\cdots x_00^\omega$ such that $d(x)=x_kx_{k-1}\cdots x_1x_0\bullet 0^\omega$. The set of $(-\beta)$-integers is then defined as \[\mathbb{Z}_{-\beta}=\{x\in\mathbb{R}\ |\ x=\gamma(a_{k-1}a_{k-2}\cdots a_1 a_0), \text{ $a_{k-1}a_{k-2}\cdots a_1 a_0 0^\omega$ is $(-\beta)$-admissible}\,,\] or equivalently \[\mathbb{Z}_{-\beta}=\bigcup_{i\geq 0}(-\beta)^i T^{-i}(0)\,.\]
\end{de}

Note that $(-\beta)$-expansions of real numbers are not necessarily unique. As was said before, uniqueness holds if and only if $l\in\big(-\frac{\beta}{\beta+1},-\frac{1}{\beta+1}\big]$. Let us demonstrate this ambiguity on the following example.

\begin{ex}\label{ex:prusvih}
Let $\beta$ be the greater root of the polynomial $x^2-2x-1$, i.e. $\beta=1+\sqrt{2}$, and let $[l,l+1)=\big[-\frac{\beta^9}{\beta^9+1},\frac{1}{\beta^9+1}\big)$. Note that $[l,l+1)$ is not invariant under division by $(-\beta)$. 

If we want to find the $(-\beta)$-expansion of number $x\notin[l,l+1)$, we have to find such $k\in\mathbb{N}$ that $\frac{x}{(-\beta)^k}\in[l,l+1)$, compute $d\big(\frac{x}{(-\beta)^k}\big)$ by definition and then shift the fractional point by $k$ positions to the right. The problem is that, in general, different choices of the exponent $k$ may give different $(-\beta)$-admissible digit strings which all represent the same number $x$.

Let us find possible $(-\beta)$-expansions of $1$. It can be shown that $\frac{1}{(-\beta)^k}\in[l,l+1)$ if and only if $k\in\mathbb{N}\setminus\{0,2,4,6,8\}$ and there are $5$ $(-\beta)$-admissible digit strings representing $1$, computed from $(-\beta)$-expansions of $\frac{1}{(-\beta)^k}$ for $k=1,3,5,7,9$ respectively: \[1\bullet 0^\omega\quad=\quad 120\bullet 0^\omega\quad=\quad 13210\bullet 0^\omega\quad=\quad 1322210\bullet 0^\omega\quad=\quad 132222210\bullet 0^\omega\,.\]
\end{ex}

Let us mention some straightforward observations on the properties of $\mathbb{Z}_{-\beta}$:

\begin{itemize}
\item $\mathbb{Z}_{-\beta}$ is nonempty if and only if $0\in\mathcal{A}_{-\beta,l}$, i.e. if and only if $l\in(-1,0]$.
\item The definition implies $-\beta\mathbb{Z}_{-\beta}\subset\mathbb{Z}_{-\beta}$.
\item A phenomenon unseen in R\'enyi numeration arises, there are cases when the set of $(-\beta)$-integers is trivial, i.e. when $\mathbb{Z}_{-\beta}=\{0\}$. This happens if and only if both numbers $\frac{1}{\beta}$ and $-\frac{1}{\beta}$ are outside of the interval $[l,l+1)$. This can be reformulated as \[\mathbb{Z}_{-\beta}=\{0\}\quad\Leftrightarrow\quad \beta<-\frac{1}{l}\ \text{ and }\ \beta\leq\frac{1}{l+1}\,,\] and it can be seen that the strictest limitation for $\beta$ arises when $l=-\frac{1}{2}$. This implies for any choice of $l\in\mathbb{R}$:\[\mathbb{Z}_{-\beta}\neq\emptyset\ \text{ and }\ \beta\geq 2\quad\Rightarrow\quad\mathbb{Z}_{-\beta}\supsetneq\{0\}\,.\]
\item It holds that $\mathbb{Z}_{-\beta}=\mathbb{Z}$ if and only if $\beta\in\mathbb{N}$.
\end{itemize}

\begin{pozn}
As was shown in Example~\ref{ex:prusvih}, in a completely general case of $(-\beta)$-expansions, there is a problem with ambiguity. Because of this, in the following we shall limit ourselves to the choice $l\in\big[-\frac{\beta}{\beta+1},-\frac{1}{\beta+1}\big]$. Note that we allow Ito-Sadahiro case $l=-\frac{\beta}{\beta+1}$, which also contains ambiguities, but only in countably many cases, which can be avoided by introducing a notion of strong $(-\beta)$-admissibility.
\end{pozn}

\begin{de}
Let $x_1x_2x_3\cdots\in\mathcal{A}_{-\beta,l}$. We say that \[x_1x_2x_3\cdots\text{ is strongly $(-\beta)$-admissible}\quad\text{ if }\quad 0x_1x_2x_3\cdots\text{ is $(-\beta)$-admissible}\,.\]
\end{de}

\begin{pozn}
Note that if $l\in\big(-\frac{\beta}{\beta+1},-\frac{1}{\beta+1}\big]$, the notions of strong admissibility and admissibility coincide. In the case $l=-\frac{\beta}{\beta+1}$, the only numbers with non-unique expansions are those of the form $(-\beta)^kl$, which have exactly two possible expansions using digit strings $l_1l_2l_3\cdots$ and $1l_1l_2l_3\cdots$. While both are $(-\beta)$-admissible, only the latter is also strongly $(-\beta)$-admissible.
\end{pozn}

In order to describe distances between adjacent $(-\beta)$-integers, we will study
ordering of finite digit strings in the alternate order. Denote by $\mathcal{S}(k)$
the set of infinite $(-\beta)$-admissible digit strings such that erasing a prefix
of length $k$ yields $0^\omega$, i.e. for $k\geq 0$, we have
\[
\mathcal{S}(k)=\{a_{k-1}a_{k-2}\cdots a_00^\omega \mid
a_{k-1}a_{k-2}\cdots a_00^\omega \text{ is $(-\beta)$-admissible}\}\,,
\]
in particular $\mathcal{S}(0) = \{0^\omega\}$. For a fixed $k$, the
set  $\mathcal{S}(k)$ is finite. Denote by $\mathrm{Max}(k)$ the string
$a_{k-1}a_{k-2}\cdots a_00^\omega$ which is maximal in $\mathcal{S}(k)$
with respect to the alternate order and by $\max(k)$ its
prefix of length $k$, i.e. $\mathrm{Max}(k)  = \max(k)0^\omega$.
Similarly, we define $\mathrm{Min}(k)$ and {\bf  $\min(k)$}. Thus,
\[
\mathrm{Min}(k) \preceq_{alt} r \preceq_{alt} \mathrm{Max}(k)\,,
\qquad\text{ for all digit strings $r \in \mathcal{S}(k)$.}
\]

With this notation we can give a theorem describing distances in $\mathbb{Z}_{-\beta}$ valid for cases $l\in\big[-\frac{\beta}{\beta+1},-\frac{1}{\beta+1}\big]$. Note that for case $l=-\frac{\beta}{\beta+1}$ it was proven in $\cite{ADMP}$.

\begin{thm}\label{prop:candidates_for_distances} Let $x <y$ be two consecutive
$(-\beta)$-integers. Then there exist a finite string $w$ over the alphabet
$\mathcal{A}_{-\beta,l}$,  a non-negative integer
$k\in\{0,1,2,\dots\}$ and a positive digit $d \in
\mathcal{A}_{-\beta,l}\setminus \{0\}$ such that
$w(d-1)\mathrm{Max}(k)$ and $wd\mathrm{Min}(k)$ are strongly $(-\beta)$-admissible strings and
$$
\begin{array}{lcll}
x = \gamma(w(d-1)\max(k)) &<&  y =\gamma(wd\min(k))\quad &\hbox{for $k$ even},\\
x = \gamma(wd\min(k)) &<&  y = \gamma(w(d-1)\max(k)) \quad &\hbox{for $k$ odd}.
\end{array}
$$
In particular,  the distance $y-x$  between these
$(-\beta)$-integers depends only on $k$ and  equals to
\begin{equation}\label{Delty}
\Delta_k:=\Big|(-\beta)^k + \gamma\big(\mathrm{min}(k)\big) -
  \gamma\big(\mathrm{max}(k)\big)\Big|\,.
  \end{equation}
\end{thm}

\section{Coding $\mathbb{Z}_{-\beta}$ by an infinite word}

Note that in order to get an explicit formula for distances from Theorem~\ref{thm:hlavni}, knowledge of reference strings $\min(k)$ and $\max(k)$ is necessary. These depend on both reference strings $d(l)$ and $d^*(l+1)$. Concerning the form of $\min(k)$ and $\max(k)$ we provide the following proposition.

\begin{prop}\label{prop_tvary_minmax}
Let $\beta>1$. Denote $d(l)=l_1l_2l_3\cdots$, $d^*(l+1)=r_1r_2r_3\cdots$.
\begin{itemize}
\item $\min(0)=\max(0)=\varepsilon$,
\item for $k\geq 1$ either $\min(k)=l_1l_2\cdots l_k$ or there exists $m(k)\in\{0,\cdots\!,k\!-\!1\}$ such that \[\min(k)=\left\{\begin{array}{ll}
\!l_1l_2\cdots(l_{k-m(k)}\!+\!1)\min(m(k)) & \text{if }k\!-\!m(k)\text{ even}\\
&\\
\!l_1l_2\cdots(l_{k-m(k)}\!-\!1)\max(m(k)) & \text{if }k\!-\!m(k)\text{ odd}
\end{array}\right.\]
\item for $k\geq 1$ either $\max(k)=r_1r_2\cdots r_k$ or there exists $m'(k)\in\{0,\cdots\!,k\!-\!1\}$ such that \[\max(k)=\left\{\begin{array}{ll}
\!r_1r_2\cdots(r_{k-m'(k)}\!-\!1)\max(m'(k)) & \text{if }k\!-\!m'(k)\text{ even}\\
&\\
\!r_1r_2\cdots(r_{k-m'(k)}\!+\!1)\min(m'(k)) & \text{if }k\!-\!m'(k)\text{ odd}
\end{array}\right.\]
\end{itemize}
\end{prop}

Computing $\min(k)$ and $\max(k)$ for a general choice of $l$ may lead to difficult discussion, however, in special cases an important relation between $d(l)$ and $d^*(l+1)$ arises and eases the computation. Examples were given in Remarks~\ref{pozn:IS} and \ref{pozn:bal}. 

Let us now describe how we can code the set of $(-\beta)$-integers by an infinite word over the infinite alphabet $\mathbb{N}$. 

Let $(z_n)_{n\in\mathbb{Z}}$ be a strictly increasing sequence satisfying \[z_0=0\quad\text{and}\quad\mathbb{Z}_{-\beta}=\{z_n\ |\ n\in\mathbb{Z}\}\,.\] We define a bidirectional infinite word over an infinite alphabet ${\bf v}_{-\beta} \in\mathbb{N}^\mathbb{Z}$, which codes the set of $(-\beta)$-integers. According to Theorem~\ref{prop:candidates_for_distances}, for any $n\in\mathbb{Z}$ there exist a unique $k\in\mathbb{N}$, a word $w$ with prefix $0$ and a letter $d$ such that \[z_{n+1}-z_n=\big|\gamma(w(d-1)\max(k))-\gamma(wd\min(k))\big|\,.\] We define the word ${\bf v}_{-\beta}=(v_i)_{i\in\mathbb{Z}}$ by $v_n=k$.

\begin{thm}
Let ${\bf v}_{-\beta}$ be the word associated with $(-\beta)$-integers. There exists an antimorphism $\Phi:\mathbb{N}^*\rightarrow\mathbb{N}^*$ such that $\Psi=\Phi^2$ is a non-erasing non-identical morphism and $\Psi({\bf v}_{-\beta})={\bf v}_{-\beta}$. $\Phi$ is always of the form \[\Phi(2l)=S_{2l}(2l+1)\widetilde{R_{2l}}\quad\text{and}\quad\Phi(2l+1)=R_{2l+1}(2l+2)\widetilde{S_{2l+1}}\,,\] where $\widetilde{u}$ denotes the reversal of the word $u$ and words $R_j$, $S_j$ depend only on $j$ and on $\min(k),\max(k)$ with $k\in\{j,j+1\}$.
\end{thm}

The proof is based on the self-similarity of $\mathbb{Z}_{-\beta}$, i.e. $-\beta\mathbb{Z}_{-\beta}\subset\mathbb{Z}_{-\beta}$, and on the following idea. Let $x=\gamma(w(d-1)\max(k)) < y=\gamma(wd\min(k))$ be two neighbours in $\mathbb{Z}_{-\beta}$ with gap $\Delta_k$ and suppose only $k$ even. If we multiply both $x$ and $y$ by $(-\beta)$, we get a longer gap with possibly more $(-\beta)$-integers in between. It can be shown that between $-\beta y$ and $-\beta x$ there is always a gap $\Delta_{k+1}$. Hence the description is of the form $\Phi(k)=S_{k}(k+1)\widetilde{R_{k}}$, where the word $S_k$ codes the distances between $(-\beta)$-integers in $[\gamma(wd\min(k)0), \gamma(wd\min(k+1))]$ and, similarly, $R_k$ encodes distances within the interval $[\gamma(w(d-1)\max(k)0),\gamma(w(d-1)\max(k+1))]$.

As it turns out, in some cases (mostly when reference strings $l_1l_2l_3\cdots$ and $r_1r_2r_3\cdots$ are eventually periodic of a particular form) we can find a letter-to-letter projection to a finite alphabet $\Pi:\mathbb{N}\rightarrow\mathcal{B}$ with $\mathcal{B}\subset\mathbb{N}$, such that ${\bf u}_{-\beta}=\Pi{\bf v}_{-\beta}$ also encodes $\mathbb{Z}_{-\beta}$ and it is a fixed point of a an antimorphism $\varphi=\Pi\circ\Phi$ over the finite alphabet $\mathcal{B}$. Clearly, the square of $\varphi$ is then a non-erasing morphism over $\mathcal{B}$ which fixes ${\bf u}_{-\beta}$.

Let us mention that $(-\beta)$-integers in the Ito-Sadahiro case $l=-\frac{\beta}{\beta+1}$ are also subject of \cite{Wolfgang}. For $\beta$ with eventually periodic $d(l)$, Steiner finds a coding of $\mathbb{Z}_{-\beta}$ by a finite alphabet and shows, using only the properties of the $(-\beta)$-transformation, that the word is a fixed point of a non-trivial morphism. Our approach is of a combinatorial nature, follows a similar idea as in \cite{ADMP} and shows existence of an antimorphism for any base $\beta$.

To illustrate the results, let us conclude this contribution by an example.

\begin{ex}
Let $\beta$ be the real root of $x^3-3x^2-4x-2$ ($\beta$ Pisot, $\approx 4.3$) and $l=-\frac{1}{2}$. The admissibility condition gives us for any admissible digit string $(x_i)_{i\geq 0}$: \[201^\omega\preceq_{alt}x_ix_{i+1}x_{i+2}\cdots\prec_{alt}\overline{2}0\overline{1}^\omega\,\quad\text{for all } x\geq 0\,.\]

We obtain \[\min(0)=\varepsilon, \quad\min(1)=2, \quad\min(2)=20\] and \[\min(2k+1)=20(11)^{k-1}0,\quad \min(2k+2)=20(11)^k \quad\text{ for $k\geq 1$}\,.\] Clearly it holds that $\max(i)=\overline{\min(i)}$ for all $i\in\mathbb{N}$.

Theorem~\ref{prop:candidates_for_distances} gives us the following distances within $\mathbb{Z}_{-\beta}$: \[\Delta_0=1, \quad\Delta_1=-1+\frac{4}{\beta}+\frac{2}{\beta^2}, \quad\text{ and }\quad\Delta_{2k}=1-\frac{2}{\beta}-\frac{2}{\beta^2},\quad\Delta_{2k+1}=1+\frac{2}{\beta}+\frac{2}{\beta^2}\quad\text{ for $k\geq 1$}\,.\]

Finally, the antimorphism $\Phi:\mathbb{N}^*\rightarrow\mathbb{N}^*$ is given by \begin{align*}
0 \rightarrow&\ 0^210^2\,,\\
1 \rightarrow&\ 2\,,\\
2 \rightarrow&\ 3\,,\\
\intertext{and for $k\geq 1$}
2k+1 \rightarrow&\ 0^210(2k+2)010^2\,,\\
2k+2 \rightarrow&\ 2k+3\,.
\end{align*}
It can be easily seen that a projection from $\mathbb{N}$ to a finite alphabet exists and a final antimorphism $\varphi:\{0,1,2,3\}^*\rightarrow\{0,1,2,3\}^*$ is of the form \begin{align*}
0 \rightarrow&\ 0^210^2,\\
1 \rightarrow&\ 2,\\
2 \rightarrow&\ 3,\\
3 \rightarrow&\ 0^2102010^2.
\end{align*}
\end{ex}

\providecommand{\urlalt}[2]{\href{#1}{#2}}
\providecommand{\doi}[1]{doi:\urlalt{http://dx.doi.org/#1}{#1}}
\renewcommand{\refname}{Bibliography}

\end{document}